\documentclass[reprint, twocolumn, amssymb, amsmath, nobibnotes, superscriptaddress, prl]{revtex4-2}
\usepackage{graphicx}
\usepackage{float}
\usepackage{dcolumn}
\usepackage{bm}
\usepackage[dvipsnames]{xcolor}
\usepackage[colorlinks=true]{hyperref}
\hypersetup{
    colorlinks=true,       
    linkcolor=blue,     
    citecolor=blue,    
    urlcolor=blue      
} 
\usepackage{natbib}
\usepackage{cleveref}
\usepackage[caption=false]{subfig}
\usepackage{wasysym}

\begin{document}

\title{Many-body localization enables iterative quantum optimization}

\author{Hanteng Wang}
\email{wanghanteng@sjtu.edu.cn}
\affiliation{%
 School of Physics and Astronomy, University of Minnesota, Minneapolis, Minnesota 55455, USA
}%
\affiliation{%
 Shanghai Center for Complex Physics, School of Physics and Astronomy, Shanghai Jiao Tong University, Shanghai 200240, China
}%

\author{Hsiu-Chung Yeh}
\affiliation{%
 School of Physics and Astronomy, University of Minnesota, Minneapolis, Minnesota 55455, USA
}%

\author{Alex Kamenev}
\email{kamenev@physics.umn.edu}

\affiliation{%
 School of Physics and Astronomy, University of Minnesota, Minneapolis, Minnesota 55455, USA
}%
\affiliation{%
 William I. Fine Theoretical Physics Institute, University of Minnesota, Minneapolis, Minnesota 55455, USA
}%

\date{\today}

\begin{abstract}
We suggest an iterative quantum  protocol, allowing to solve optimization problems with a 
glassy energy landscape. It is based on a periodic cycling around the tricritical point of the many-body localization transition. This ensures that each iteration leads to a non-exponentially small probability to find a lower local energy minimum. The other key ingredient is to tailor the cycle parameters to a currently achieved optimal state (the ``reference'' state) and to reset them once a deeper minimum is found. We show that, if the position of the tricritical point is known, 
the algorithm allows to approach the absolute minimum with any given precision in a polynomial time.      
\end{abstract}

\maketitle


Optimization problems are ubiquitous \cite{garey1979computers,arora2009computational}. A large subclass of them is  discrete optimization tasks, which may be mapped onto spin models with the optimal solution being a ground state of a certain classical spin Hamiltonian.  The  optimization problems are hard due to  the  spin-glass phase \cite{barahona1982computational,lucas2014ising,mezard1987spin}, i.e. presence of  multiple local minima in the energy landscape of the corresponding model. The idea of utilizing quantum tunneling in order to facilitate transitions between these local minima was coined a long time ago. Probably the earliest and most transparent way of doing it is realized via the adiabatic  quantum annealing (QA) procedure \cite{kadowaki1998quantum,farhi2000quantum,farhi2001quantum}. 
Its bottleneck is associated with exponentially small energy gaps between instantaneous energy levels of the corresponding {\em quantum} Hamiltonian \cite{amin2009first,altshuler2010anderson,jorg2010energy,jorg2008simple,knysh2010relevance,knysh2016zero,young2008size}. 
Those lead to  Landau-Zener transitions \cite{zener1932non,sinitsyn2002multiparticle,sinitsyn2016solvable}, which take the system out of its adiabatic ground state.  As a result in order to succeed, the  QA should be performed exponentially slow. 

This stimulates interest in constructing approximate  {\em diabatic} protocols \cite{albash2018adiabatic,crosson2021prospects},  collectively known as quantum approximate optimization algorithms \cite{farhi2014quantum,farhi2016quantum,wang2018quantum,zhou2020quantum}. The idea is to force the system to gradually approach its GS with a relatively fast running cycles \cite{ohkuwa2018reverse,yamashiro2019dynamics,passarelli2020reverse,king2018observation}.

Here we suggest an iterative  quantum algorithm which runs along a closed cycle in a space of parameters. The key 
observation is that the cycle must encircle a tricritical point of the many-body localization (MBL) \cite{altshuler1997quasiparticle,basko2006metal,gornyi2005interacting,oganesyan2007localization,pal2010many,laumann2014many,baldwin2017clustering,mukherjee2018many,laumann2012quantum} transition. 
The three phases coming together at the tricritical point are the spin-glass, the MBL paramagnet and the delocalized paramagnet.  The cycle starts in the spin-glass and goes successively into MBL and delocalized paramagnets before 
returning back to the spin-glass, where the projective measurement is performed. We show that iterations of such cycle
lead to a systematic decrease of energy of the measured state. The cycle duration and a number of required cycles scale 
algebraically with the system size. Given a desired precision of the optimization, the cycle trajectory should pass increasingly close to the tricritical point.     

\begin{figure}[t]
  \centering
  \includegraphics[width=0.4\textwidth]{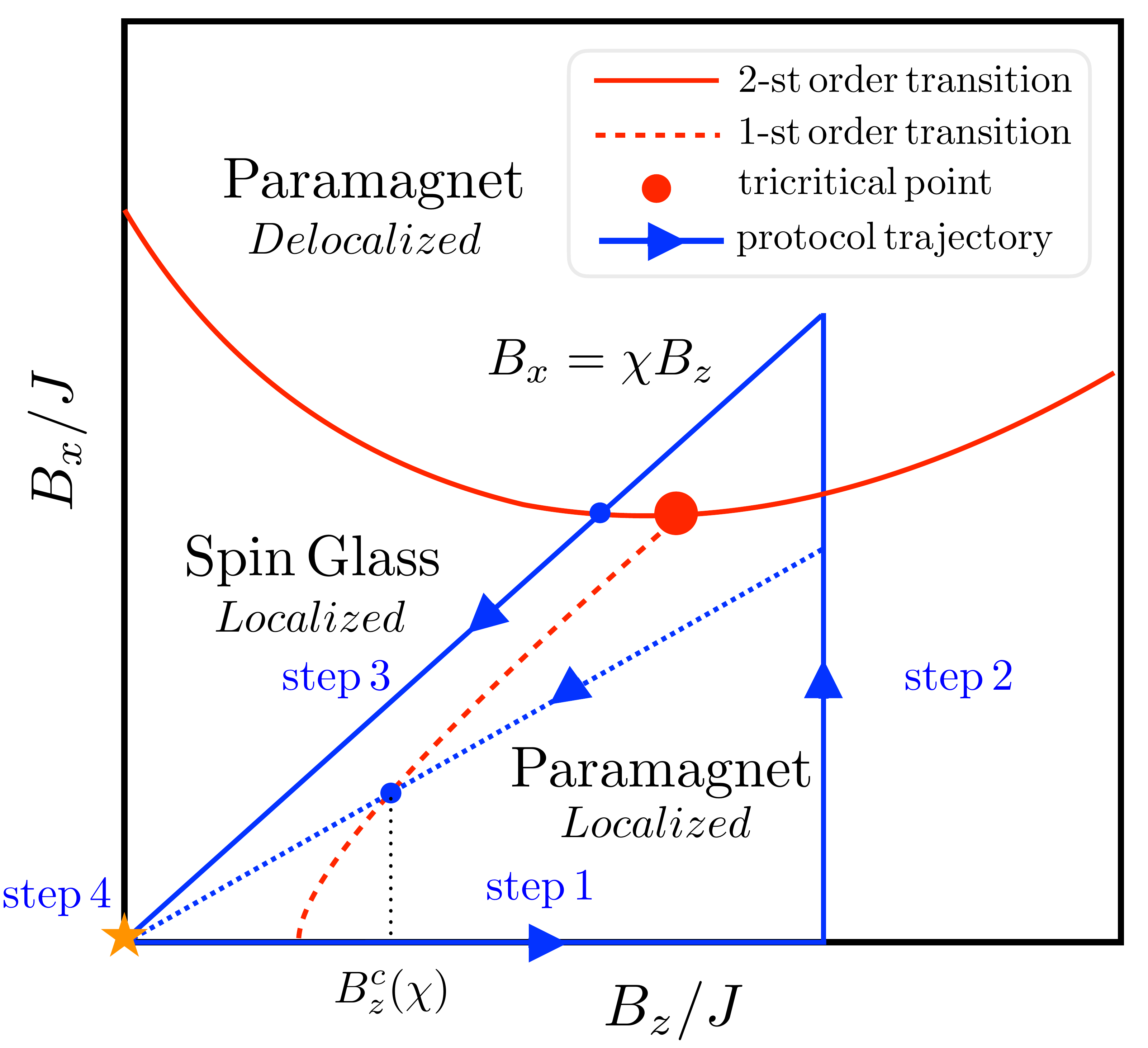}
  \caption{ {\bf Phase diagram and the protocol.}  Phase diagram of the Hamiltonian (\ref{eq:full_Hamiltonian}) for a specific reference state. Full red line indicates  a   2-nd order transition \cite{young2017stability} between MBL and delocalized paramagnet; the dashed red line is the 1-st order transition between the glass and MBL paramagnet. They meet at the tricritical point. The blue lines with arrows represent one optimization cycle. The dotted blue line is a subcritical slope, $\chi<\chi_c$, which most likely brings the system back to the initial reference state.}
  \label{fig:Protocol}
\end{figure}

The iterative optimization has already appeared in the literature, see e.g.   Refs.~[\onlinecite{king2018observation,ohkuwa2018reverse,yamashiro2019dynamics,passarelli2020reverse}]. We found it 
useful to combine it with the idea of the reference Hamiltonian \cite{perdomo2011study,chancellor2017modernizing,denchev2019quantum}. The latter calls for using a control parameter (eg. a longitudinal magnetic field) which is collinear with a local Bloch sphere direction of the individual qubits.  This leads to 
an adjustment of the cycle parameters according to a result of the measurement taken at the end of the previous cycle.    
We show that such strategy allows to navigate the system arbitrarily close to the MBL tricritical point, as required 
by the proposed algorithm. As an example of  optimization in a spin-glass system, we use  Sherrington-Kirkpatrick (SK) model \cite{sherrington1975solvable}, whose  MBL properties are discussed Refs.~[\onlinecite{laumann2014many,baldwin2017clustering,mukherjee2018many,laumann2012quantum}].  \\

\noindent {\bf Results}

\noindent {\bf Iterative quantum optimization  protocol.} As an example of an optimization problem with a glassy landscape we choose a realization of the 
Sherrington-Kirkpatrick (SK) model \cite{sherrington1975solvable} specified by a Hamiltonian 
\begin{equation} \label{eq:SK}
  H_{\mathrm{SK}}= \sum_{ij}^N J_{ij} \sigma^z_i \sigma^z_j. 
\end{equation}
Here $\sigma^z_i$ are $z$-Pauli matrices, which represent binary optimization parameters, labeled  by $i=1,2,\ldots N$. The cost function is chosen to be quadratic in these parameters given by a cost matrix, $J_{ij}$. In our examples its matrix elements are taken from independent  Gaussian distributions with zero mean and variance $J^2/N$.  Eigenstates of the Hamiltonian, denoted as $\alpha=1,2,\ldots, 2^N$, are encoded by bit-strings, $\{ s_i^{\alpha} \}$, with $s_i^\alpha =\pm 1$ showing ``up'' or ``down'' polarization of the $i$-th spin.   The corresponding eigenenergies are  $E_{\alpha}= \sum_{ij}^N J_{ij} s^\alpha_i s^\alpha_j$. Since all the terms in the Hamiltonian (\ref{eq:SK}) commute with each other, the problem is purely classical.

It is known \cite{mezard1987spin,crisanti2003complexity,cavagna2004numerical} that  $E_{\alpha}$ form a glassy landscape with exponentially many local minima (i.e. states such that flipping {\em any} one (or even a few) spins results in energy being increased). Simulated classical annealing is typically trapped into one of such local minima. The local minima are separated from each other by  the {\em Hamming distance} of the order $N$ spin flips.  
The goal of the optimization is to find progressively deeper local minima, eventually hitting the  global one.

The conventional adiabatic QA procedure calls for modifying the Hamiltonian (\ref{eq:SK}) to add non-commutative (aka quantum) terms. The simplest of such quantum terms is (in general  time-dependent) magnetic field applied in the $x$-direction: 
\begin{equation}
  H(t)= H_{\mathrm{SK}}  + B_x(t) H_{\mathrm{q}},\qquad \ H_{\mathrm{q}}=-\sum_{i=1}^N\sigma_i^x.
\end{equation}

If the $x$-magnetic field is initiated to be  large, $B_x \gg J$, the ground state is close to all spins being polarized in the $x$-direction  \cite{laumann2014many,baldwin2017clustering}. Such ground state is separated by a large gap, $\sim B_x$, from the rest of the spectrum. Cooling the system down to a temperature $T\ll B_x$ puts it almost surely in its true ground state. One then slowly decreases 
$B_x(t)$ down to zero so that the Hamiltonian goes back to the pure SK model (\ref{eq:SK}). If this process is adiabatic, the state of the system follows its instanteneous ground state and arrives at the global SK  minimum. 
For the system to {\em not} undergo any Landau-Zener transition, the annealing rate should be $\tau^{-1}_{\mathrm{anneal}} \ll  \Delta_{\mathrm{min}}^2/B_x$, where $\Delta_{\mathrm{min}}$ is a minimal avoiding crossing gap, encountered by the ground state. As argued in Refs.~\cite{amin2009first,altshuler2010anderson,jorg2010energy} some of these gaps are exponentially small, demanding an exponentially long annealing time,  $\tau_{\mathrm{anneal}}$.

Hereby we suggest an iterative  cyclic algorithm capable of  systematically approaching the ground state, while not  being exponentially slow. 
Before the first cycle starts one performs a simulated classical annealing, arriving at one of the local minima, which we'll call a reference state, $\{s_i^{r}\}$.    Each cycle consists of the four successive  steps summarized in Fig.~\ref{fig:Protocol}:

{\em Step 1.}  The qubit array is initialized to the reference state and is programed to represent the following Hamiltonian 
\begin{equation} \label{eq:full_Hamiltonian}
  H(t)= H_{\mathrm{SK}}  + B_x(t) H_{\mathrm{q}}+ B_z(t) H_{\mathrm{ref}}^r,
\end{equation}
where the $z$-field in the reference  Hamiltonian is tailor-made to be co-directed with all the spins of the given reference bit-string,  $\{s_i^{r}\}$, 
\begin{equation} \label{eq: ref Hamiltonian}
  H_{\mathrm{ref}}^r= - \sum_{i} s_i^{r} \sigma_i^z.
\end{equation}
One starts from the pure SK model, $B_x=B_z=0$, and then increases $B_z(t)$ from zero passing the critical field $B_z^c$, separating the spin-glass phase from the paramagnet. Since $B_x=0$   in step 1, the Hamiltonian is purely classical and the system remains in the reference state, no matter how fast  $B_z$ is increased. In fact, all the states remain to be pure bit-strings of $H_{\mathrm{SK}}$, but their relative energies do change. The $H_{\mathrm{ref}}^r$ is chosen in a way to push the energy of the reference state sharply  down: $E_r(B_z)=E_r(0) - NB_z$. The other local 
minima are far in the Hamming distance from the reference state  and thus evolve typically as   
$E_\alpha(B_z)=E_\alpha(0) \pm \sqrt{N} B_z$. As a result, soon enough the reference state is the unique ground state, separated by the gap. 
This first happens  at the critical field $B_z^c$.  

{\em Step 2}: $B_x$ is increased while $B_z$ is fixed. The gap in the paramagnetic phase is proportional to the total magnetic field $\sqrt{B_x^2+B_z^2}$,  and is independent of the system size. One does not need an exponential or even a power law (in system size) long time to increase $B_x$ while keeping the system in the ground state of the full Hamiltonian (\ref{eq:full_Hamiltonian}). However, since the full Hamiltonian is now quantum, its ground state is a superposition of many 
bit-string states. The $B_x$ is increased until it reaches a certain ratio with the $z$-field: $\chi = B_x/B_z$.

{\em Step 3}: Decreasing $B_z$ and $B_x$ keeping the fixed ratio $\chi$ between them. Along this path the system again crosses the phase boundary between the paramagnetic and  the glassy phases. This boundary is marked by the first avoiding crossing transition between the ground state and the lowest excited state.  The size of the corresponding gap strongly depends on the slope $\chi$, which we discuss in detail in the next section. The upshot is that Landau-Zener transitions may occur during this part of the cycle, but with an overwhelming  probability they leave the system in a state with an energy, which is lower than that of the initial reference state.  The main danger is that the system remains in the reference state. This may be avoided, however, by a careful choice of the slope $\chi$.  

{\em Step 4.}  After both $B_x$ and $B_z$ reach zero in the end of the step 3, the system ends up in  a superposition state. 
Now the measurement of each qubit is performed and the state collapses to a certain bit-string. Starting from this measured bit-string, the simulated annealing leads the system down to a nearest local minimum. If the energy of this  new local minimum is less than that of the reference state, it is taken as the new reference state and the cycle is repeated  from step 1. If, however, 
its energy is larger or the same, the system is initiated back to the old reference state and the cycle is again repeated from step 1.

Three key features of this protocol qualitatively improve its performance vis-a-vis the conventional QA. First, the reference state is iteratively set to be the minimal energy local minimum found in all previous trials. This way the reference energy  
never increases.  Second, the choice of the reference Hamiltonian guarantees that  Zener transitions in step 3  almost 
always decrease the energy. Third (and most significant), cycling around the tricritical point of the MBL transition 
allows to accomplish such energy decrease in a polynomial time. The second and the third items on this list are explained in the next section. \\

\begin{figure}[t]
  \centering
  \includegraphics[width=0.48\textwidth]{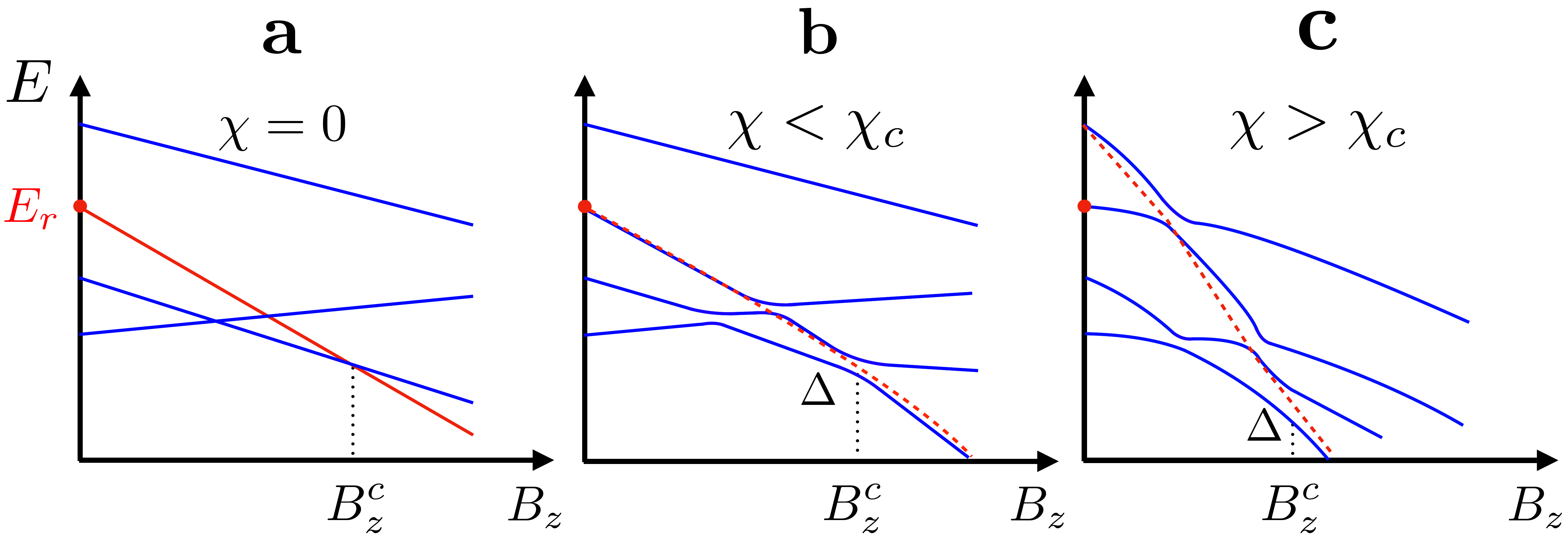}
  \caption{ {\bf Spectrum and Landau-Zener transition.} A sketch of the spectrum of Eq.~(\ref{eq:full_Hamiltonian}) for different values of the slope $\chi=B_x/B_z$. The red dots represent the SK energy of the reference state. In {\bf a}, the red line depicts the energy of reference state. In {\bf b} and {\bf c}, the reference state is the eigenstate only at $B_z = B_x = 0$. The red dashed line indicates a {\em diabatic} trajectory of step 3, undergoing  Landau-Zener transition (from large $B_z$ to small $B_z$).
  }
  \label{fig:Sketch of Spectrum}
\end{figure}

\noindent {\bf MBL transition and the phase diagram.} To illustrate the statements made above, consider Fig.~\ref{fig:Sketch of Spectrum} depicting schematically the adiabatic spectrum of the Hamiltonian (\ref{eq:full_Hamiltonian}) vs. $B_z$ for  several fixed slopes $\chi$, such that  $B_x=\chi B_z$.   Figure~\ref{fig:Sketch of Spectrum}a shows $\chi=0$ case, which corresponds to the step 1 of the protocol. Since $B_x=0$, the Hamiltonian is classical and there are no transitions between the states. The corresponding energy levels cross each other. The reference Hamiltonian (\ref{eq: ref Hamiltonian}) is chosen in a way to ensure that the reference state (red line in Fig.~\ref{fig:Sketch of Spectrum}a) goes down with a maximal slope. As a result, the reference state is destined to become a ground state at a certain critical field $B_z^c$. For $B_z > B_z^c$, there is a finite energy gap between the ground reference state and the rest of the spectrum. We thus refer to this phase as the paramagnet. 
Since all the states of such a paramagnet are represented by pure bit-strings, they  are perfectly many-body localized in the bit-string basis.  Notice that within the glassy phase, $B_z < B_z^c$, the reference state crosses {\em only} the states whose SK energy is less than $E_r$. 

Figure~\ref{fig:Sketch of Spectrum}b shows the spectrum for  $\chi < \chi_c$. Due to the presence of $H_\mathrm{q}$, spin flips are allowed leading to avoiding crossings gaps in the spectrum. At small $B_x$, these  gaps are exponentially small, because of  typically large (order $N$) Hamming distance between  low-energy local minima. This makes the critical field $B_z^c(\chi)$ to be well defined in the large $N$ limit. It marks the {\em first order} transition 
between MBL glass and MBL paramagnet phases. If the step 3 of the protocol is run (right to left) along this trajectory (dotted blue line in Fig.~\ref{fig:Protocol}) with a non-exponentially small  rate, the state of the system most likely follows the dashed red line. This brings  
the system back to the initial reference state, making the protocol fail.  Notice, however, that in rare cases when the state does follow the adiabatic trajectories, the energy of the system is bound to be {\em below} the initial  energy, $E_r$.

To increase the probability of adiabatic transitions lowering the energy, the gaps need to be increased. This is achieved by 
working at $\chi > \chi_c$,   Fig.~\ref{fig:Sketch of Spectrum}c. Such a strategy comes with a steep prize, however. Indeed, 
the reference state may also hybridize now with higher energy states.  This leads  to undesirable transitions increasing the energy (dashed red line in Fig.~\ref{fig:Sketch of Spectrum}c). The question is if one can benefit from energy decreasing adiabatic trajectories, without being handicapped by Zener transitions to higher energy states (the latter phenomenon 
is responsible for the failure of the conventional QA \cite{amin2009first,altshuler2010anderson,jorg2010energy,knysh2016zero,young2008size}).

To answer this question one needs to examine MBL transition on the phase diagram of our protocol, Fig.~\ref{fig:Protocol}. 
Being defined by the Hamiltonian (\ref{eq:full_Hamiltonian}), the latter is tight to a specific reference state. Depending on 
the quantum component $B_x$, this state and its neighbors may be either localized (small $B_x$) or delocalized (large $B_x$) in the bit-string basis. The transition between the two is characterized by a divergent localization-Hamming-length in the many-body Fock (i.e. bit-string) space \cite{altshuler1997quasiparticle,gornyi2005interacting,basko2006metal,oganesyan2007localization,pal2010many}. Therefore the MBL transition is of the 2-nd order \cite{laumann2014many,baldwin2017clustering,mukherjee2018many,young2017stability}. It  divides the phase space, Fig.~\ref{fig:Protocol}, onto the two disconnected regions. As explained above,  there is also the 1-st order transition between  gapless (in the large $N$ limit) spin glass phase and the gaped paramagnet, both within the localized phase. 
The latter transition is not associated with a divergent Hamming distance.  It is reasonable to expect that  the 1-st order transition line terminates at a {\em tricritical} point somewhere along the MBL transition boundary, Fig.~\ref{fig:Protocol}. 

Position of the tricritical point defines a critical slope $\chi_c$ of the step 3 part of the cycle.  For $\chi< \chi_c$ the step 3 encounters the 1-st order transition within the MBL phase. Since all states below the reference one are many-body localized, the avoiding crossing gaps are exponentially small. Unless performed adiabatically (i.e. within exponentially long time), the step 3 is bound to bring the system back to its initial reference state. 

The situation is qualitatively different for $\chi>\chi_c$. Here the step 3 trajectory passes through the second order transition between a delocalized paramagnet and localized glass phase. There is a divergent localization-Hamming-length on  the glass side.  This makes the states to be spanned by progressively wider superpositions of  bit-strings and thus leads to non-exponential gaps near the MBL transition. Indeed, from generic finite size scaling considerations of the 2-nd order transition, one expects the energy gap near the MBL transition to scale as 
\begin{equation} 
          \label{eq:scaling-Delta} 
  \Delta \propto \frac{(\chi-\chi_c)^{\theta}}{N^{\nu z}},
\end{equation}
where $\theta$ and $\nu z$ are critical exponents. 
It was recently argued \cite{knysh2016zero} that in the Hopfield model (a close cousin of SK) $\nu z=1/3$.    

Therefore if the step 3 is performed within the power-law time, $\tau_3 \sim N^{2\nu z}$, it results in a certain number of the avoiding crossing transitions taking the adiabatic turn.  Those typically happen at a relatively large 
$B_x\approx \chi_c B_z^c(\chi_c)$, close to the MBL transition.   What remains to be shown is that these transitions indeed lead to a systematic energy decrease, not overshadowed by  transitions to the higher energy states, as in  Fig.~\ref{fig:Sketch of Spectrum}c.  The key insight is that this may be achieved by tuning the slope $\chi$ closer to the critical one from above, 
$\chi \to \chi_c^+$.      

\begin{figure}[t]
  \centering
  \includegraphics[width=0.48\textwidth]{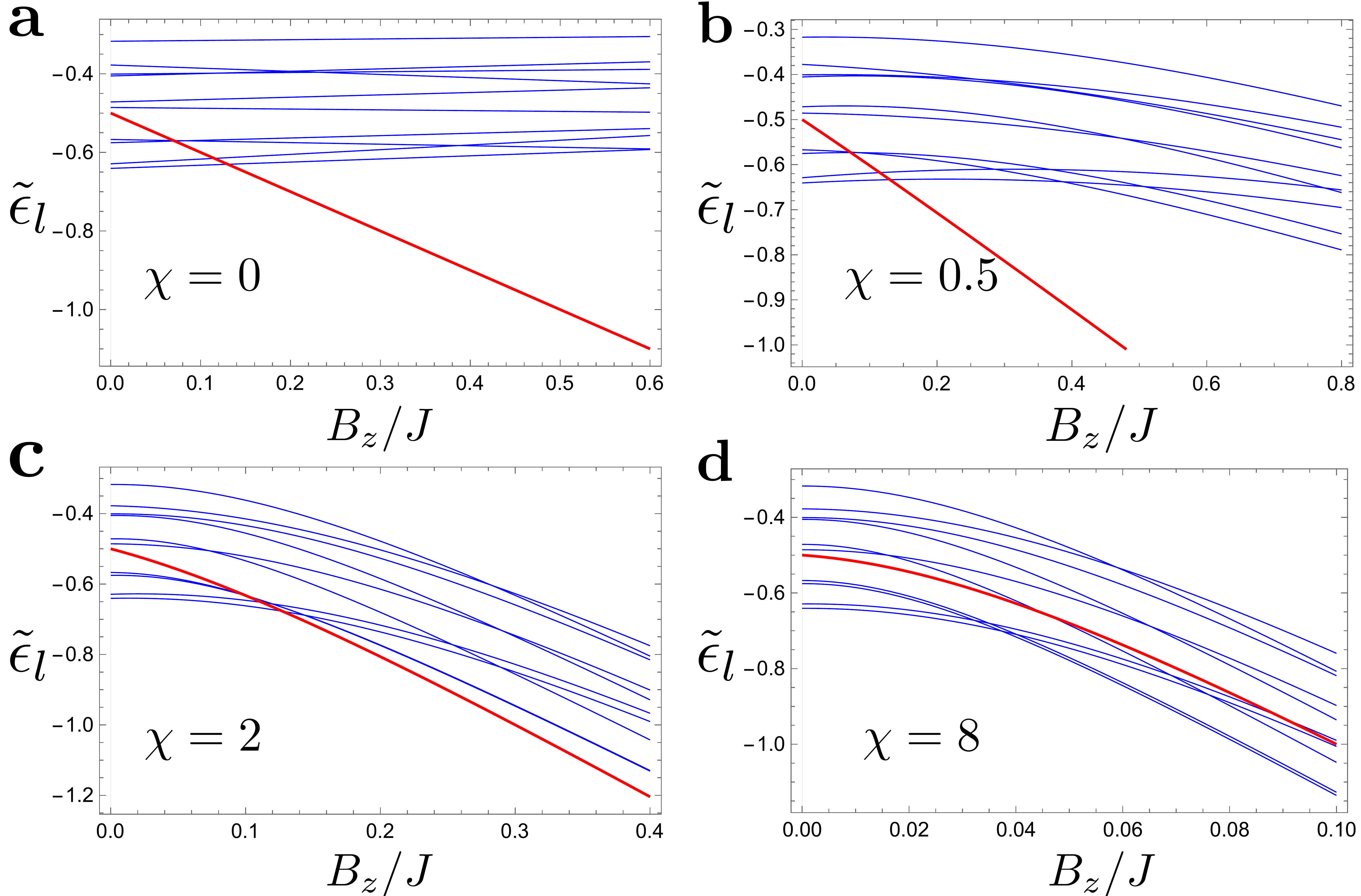}
  \caption{{\bf Spectra of isolated local minima.} Spectra of isolated local minima, shown in blue, for different $\chi$.  The reference state is shown in red. As $\chi$ increases, a progressively larger fraction 
of higher energy states curves down (due to the level repulsion from their local Hamming distance neighborhood)   to intersect the reference state.}
  \label{fig:Spectrum_X_formula}
\end{figure}

\begin{figure}[t]
  \centering
  \includegraphics[width=0.48\textwidth]{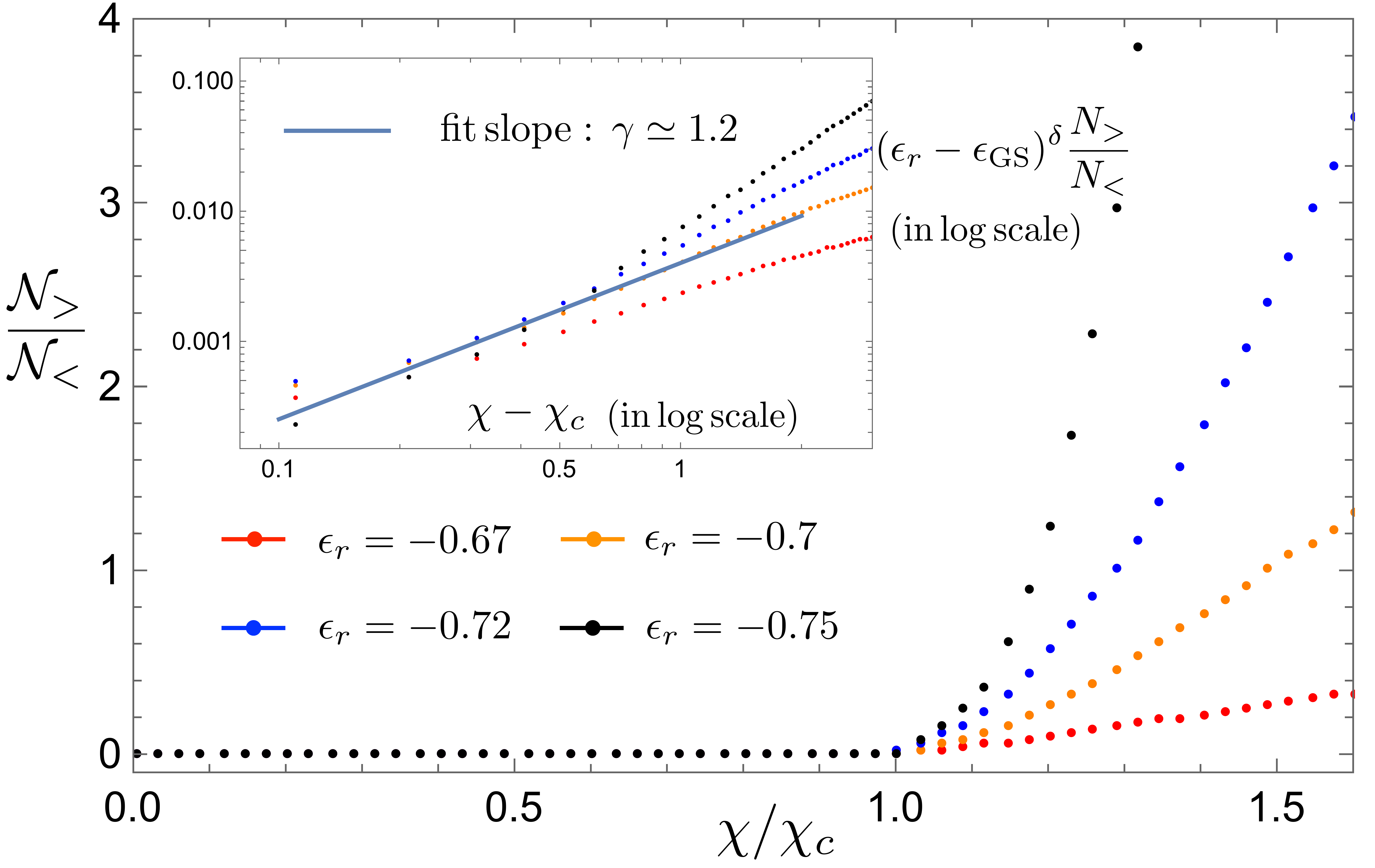}
  \caption{{\bf States ratio and scaling exponents.} Ratio of higher and lower energy isolated local minima intersecting with the reference  state as a function of the slope $\chi/\chi_c$.  The data is taken for several values of the reference state energy $\epsilon_r$. The ground state energy is 
$\epsilon_\mathrm{GS}=-0.8$.  Inset:  $(\epsilon_r - \epsilon_\mathrm{GS})^\delta \mathcal{N}_>/\mathcal{N}_<$ vs. $\chi-\chi_c$ in log-log scale;  solid line is fitting with Eq.~(\ref{eq:scaling-ratio}).}
  \label{fig:ratio_energy}
\end{figure}

To show this we numerically isolate local minima states along with their simulated annealing basins of attraction from other local minima basins. One may diagonalize Hamiltonian (\ref{eq:full_Hamiltonian}) in each of such basins (details of this procedure are described in Methods section).
This way we keep the geometry of the levels, undisturbed by the avoiding crossings generated by tunneling between  the local minima. It allows us to track exact identities of all local minima, in particular the reference state.  Figure~\ref{fig:Spectrum_X_formula} shows energies of such isolated local minima vs. $B_z$.  One can now calculate the number of local minima, with both higher energy, $\mathcal{N}_>$, and lower energy, $\mathcal{N}_<$,  crossing the reference state.     Figure \ref{fig:ratio_energy} shows the ratio $\mathcal{N}_>/\mathcal{N}_<$ vs. slope $\chi$.   As expected, for $\chi<\chi_c$ there are practically no higher energy states getting in contact with the reference one. On the other hand, the fraction of the higher energy states grows rapidly for $\chi >\chi_c$.  The smaller the energy of the reference state the faster this fraction grows. This is expected since for a deep local minimum there aren't too many other local minima below it, but there are plenty above.  
The most important lesson from Fig.~\ref{fig:ratio_energy} is what the ratio grows continuously as 
\begin{equation}
        \label{eq:scaling-ratio} 
\frac{\mathcal{N}_>}{\mathcal{N}_<} \propto \frac{(\chi-\chi_c)^\gamma}{(\epsilon_r - \epsilon_\mathrm{GS})^\delta}, 
\end{equation}
where $\gamma \approx 1.2 $ and $\delta \approx 2.0$ are critical exponents and $\epsilon_\alpha = E_\alpha/(NJ)$.  The  critical slope can depend on the reference state. In our simulations this dependence appears to be very weak, if any, with $\chi_c\approx 3.6$.

Equations (\ref{eq:scaling-Delta}) and (\ref{eq:scaling-ratio}) allow to estimate efficiency of the algorithm vis-a-vis its running time, precision and other requirements. First one fixes the desired precision, i.e. deviation from the global minimum:   $\delta_\epsilon = \epsilon_r - \epsilon_\mathrm{GS}$.  For simplicity, let us settle with the regime where every other cycle, in average, results in lowering the energy of the reference state. This amounts to the equal number of upper and lower energy local minima intersections, $\mathcal{N}_>/\mathcal{N}_<=1$. This dictates that the protocol should be run at 
$\chi-\chi_c\lesssim \delta_\epsilon^{\, \delta/\gamma}$.  Although it requires a more and more precise  knowledge of $\chi_c$, if precision is increased, the good news is that the required $\chi-\chi_c$  does not scale with the system size.   
We discuss ways of ``on the fly'' measurement of $\chi_c$ in next section.

The running time is given by a number of required cycles, $n_c$, multiplied by duration of the step 3, $\tau_3$ (steps 1, 2 and 4 are typically faster). The latter is given $\tau_3\gtrsim \Delta^{-2} \propto N^{2\nu z} \delta_\epsilon^{-2\theta\delta/\gamma}$.  Finally, assuming that every successful cycle eliminates a fraction $p<1$ of remaining lower energy states, one may estimate a number of required cycles as  $n_c\sim N/\log(1-p)$.  This leads to the total time of the optimization, which scales as 
\begin{equation} 
          \label{eq:time}
  \tau =n_c\tau_3 \propto N^{2\nu z +1}\, \delta_\epsilon^{\,-2 \theta\delta/\gamma}. 
\end{equation}
It increases with both the system size and the optimization precision, but both dependences are  power-laws, rather than exponential. 

A recent study \cite{montanari2021optimization} argued that SK model complexity scales as $N^2 C(\delta_\epsilon)$, where $C(\delta_\epsilon)$ is a function depending only on the required precision $\delta_\epsilon$. Our algorithm can't 
do better than this, even if $2\nu z < 1$. Indeed, each cycle includes classical simulated annealing step with the required time $\tau_\mathrm{sa}\sim N$.  This limits the cycle duration (but not the qubit coherence time) by $\mathrm{max}\{\tau_\mathrm{sa}, \tau_3\}$. Thus the total time scales as  $\mathrm{max}\{N^2, N^{2\nu z+1}\}$. However,  
the present quantum algorithm can match the performance of Ref.~[\onlinecite{montanari2021optimization}], if $2\nu z\leq 1$.  \\

\noindent {\bf Discussion} 

\noindent We have outlined the quantum approximate optimization algorithm, which is capable of systematically approaching the 
global minimum of a glass within the power-law (in the system size) time  (\ref{eq:time}). It is based on a variant of the quantum annealing, with the reference state specific Hamiltonian (\ref{eq:full_Hamiltonian}) and the iterative cycle encircling 
the tricritical point of the MBL transition. 

An attractive feature of the algorithm is that it does not require an exceedingly long qubit coherence time. Indeed, the projective measurement is done after every cycle. Therefore the required coherence time scales as a period of the single cycle, $\tau_3 \propto N^{2\nu z}$. Moreover, if one or a few qubits produce a faulty readout, it will be automatically corrected by simulated  classical annealing, performed after every quantum state measurement.   
Another advantage is a limited number of the required dynamical control parameters. In fact, after the Hamiltonian (\ref{eq:full_Hamiltonian}) is set, all qubits are subject to only two dynamically varying controls: $B_x(t)$ and $B_z(t)$.
There is also the measurement step, requiring a simultaneous measurement of all $\sigma^z_i$.

The main drawback of the algorithm is that it requires an exceedingly precise knowledge of the MBL trictritilal point 
slope, $\chi_c$, which is, of course, not known apriori for a given specific optimization task. Although this is 
a concern, there are ways to go about it. First, our simulations indicate that realization specific 
(and reference state specific) fluctuations of $\chi_c$ are small and decrease with increasing $N$. This means that 
$\chi_c$ can be determined (at least approximately) once for an entire broad class of optimization tasks.  Second,
the slope may be fine-tuned on the fly, while optimization cycles are running. Indeed, if the slope happens to be subcritical 
$\chi<\chi_c$, the successive iterations lead back to the same reference state. Once it happens, the slope of the next iteration needs to be somewhat increased. On the other hand, if the slope exceeds the acceptable range $\chi> \chi_c + \delta_\epsilon^{\,\delta/\gamma}$, the cycles bring the system to the local minima with higher energies. This undesirable outcome is  
corrected by decreasing the slope.   Therefore the imprecise knowledge of $\chi_c$ may be compensated by a certain 
overhead on the number of cycles, $n_c$. \\

\noindent{\bf Methods}

 \noindent {\bf Local minima isolation. } Here we discuss a phenomenological approach to numerically isolate local minima states along with their simulated annealing basins of attraction from other local minima basins.  
The low-energy Landau-Zener transitions occur only between the local minima states, $|l\rangle$, due to the fact that local minima are repelled down by their Hamming distance neighbors. To simplify the spectrum in the spin-glass phase, one may identify a basin state $\widetilde{|l\rangle}$, which is a wave packet localized at local minimum state $|l\rangle$, i.e. it is a superposition of $|l\rangle$ and its Hamming-neighbor states. Upon simulated annealing, all this states lead to  the corresponding local minimum state, i.e.  $\widetilde{|l\rangle} \rightarrow |l\rangle$. Therefore in the spin-glass phase, one can approximate the spectrum of Eq.(\ref{eq:full_Hamiltonian}) by the spectrum of local minima.

A hopping between any two local minima is typically exponentially small, since the Hamming distance is of the order of the system size $N$. An effective Hamiltonian between two basin states is
\begin{equation}
H_{l,l^{\prime}}^\mathrm{block} =
\begin{pmatrix}
\tilde{E}_l & t_{l l^{\prime}}  \\
t_{l l^{\prime}} & \tilde{E}_{l^{\prime}}
\end{pmatrix},
\end{equation}
where $t_{l l^{\prime}}$ is the effective hopping between $l$ and $l^{\prime}$ basin states with energy $\tilde{E}_l$ and $\tilde{E}_{l^{\prime}}$, which is renormalized by the Zeeman effect of $B_z$ and by repulsion from local Hamming neighborhood  due to $B_x$, i.e.
\begin{equation}
   \label{eq:energy-sigma}
\tilde{E}_l= E_l+\Sigma_l(B_z,B_x).
\end{equation}
Here $\Sigma_l(B_z,B_x)$ is the self-energy which gives the energy curves $\tilde{\epsilon}_l = \tilde{E}_l/(NJ)$ of Fig.~\ref{fig:Spectrum_X_formula} without the anti-crossing effect. 

\begin{figure}[htb]
  \centering
  \includegraphics[width=0.48\textwidth]{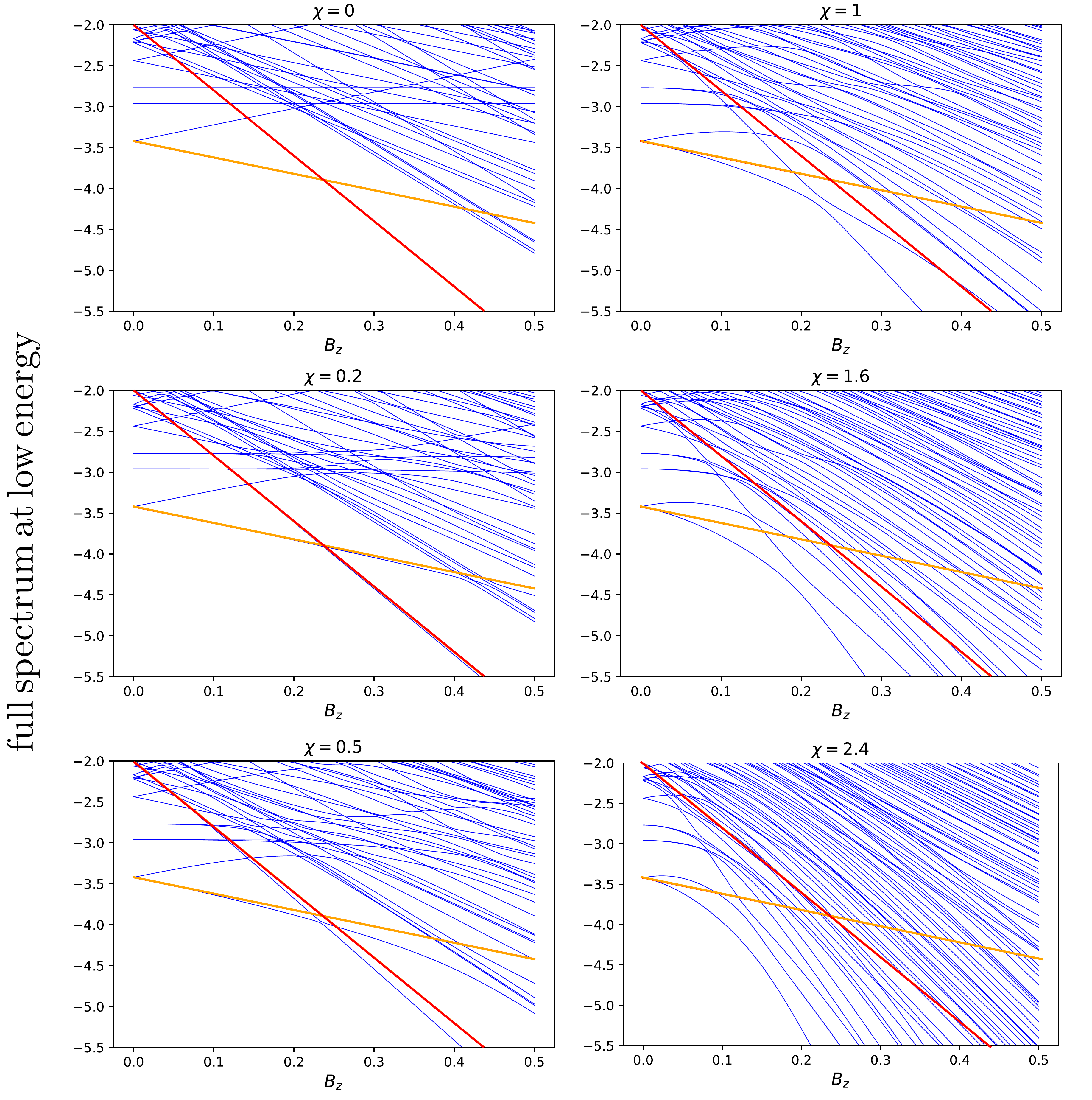}
  \caption{{\bf Exact diagonalization.} Low energy spectra of Eq.~(\ref{eq:full_Hamiltonian}) for different $\chi$. System size is  $N=10$. Red and orange straight lines are the reference and ground-state energies, correspondingly  at $\chi=0$. They are repeated on all panels as a guide for eye. When $\chi=0$, energies  are straight lines with slopes   (\ref{eq:zeeman_slope}). For larger $\chi$, energies are curved down, due to the level repulsion. Simultaneously the gaps open up. }
  \label{fig:Zeeman_z}
\end{figure}

By analyzing a small system size exact diagonalization, shown in Fig.~\ref{fig:Zeeman_z}, we found that the self-energy is 
well approximated by 
\begin{equation} \label{eq:localmin_curve}
  \Sigma_l(B_z,B_x) = NJ \left[f_l-\sqrt{(f_l + m_l B_z/J)^2 + (B_x/J)^2 }\right],
\end{equation}
where $m_l$ and $f_l$ are basin dependent phenomenological parameters discussed below. 
This expression interpolates between the limiting cases of $\chi \ll 1$ and $\chi \gg 1$. For $\chi \ll 1$ one may put  $B_x=0$,
finding $\Sigma_l = -m_l N B_z$. The corresponding slope, $m_l$, for a given local minimum $l$  measures the spin configuration overlap with the reference state $r$:
\begin{align}\label{eq:zeeman_slope}
    m_l = \frac{1}{N} \sum_{i=1}^N s_i^l \cdot s_i^r = 1-2d_l/N,
\end{align}
where $d_l$ is the Hamming distance  from the reference state to the basin $l$. We found that $d_l$'s are distributed according to a binomial distribution 
\begin{align}\label{eq:Binomial}
    P(d_{l}) = \frac{1}{2^N} {N\choose d_{l}},
\end{align}
which is natural, if one assumes totally random spin flipping (or not) to reach another local minimum.

For $\chi \gg 1$, one may start assuming $B_z = 0$.  At $B_x/J \ll 1$, Eq.~(\ref{eq:localmin_curve}) is approximate by 
\begin{align}\label{eq:Bx_curve}
    \Sigma_l \approx -NB_x^2/(2f_lJ).
\end{align}
This may be viewed as a result of the second order, in $B_x$, perturbation of the SK model.  The energy of the local minimum goes down due to the level repulsion and the second order perturbation comes from the one-spin flip states. The factor $1/(2f_l J)$  describes the average inverse energy difference between the local minimum and one-spin flip states.  The distribution of $f_l$'s is approximated by a uniform box in the interval $1/4<f_l<3/4$. 
Finally at $B_x\gg J$, the system is fully polarized with $\Sigma_l\approx -NB_x$.  Equation~(\ref{eq:localmin_curve}) is the simplest way to interpolate between all these limits, which works extremely well for small system size simulations.  

To perform larger system size simulations, leading to Fig.~\ref{fig:ratio_energy}, we statistically generate multiple local minima energy curves according to Eqs.~(\ref{eq:energy-sigma})--(\ref{eq:Bx_curve}). The distribution of SK local energies,  
$E_l$ is taken from  Refs.~[\onlinecite{crisanti2003complexity,cavagna2004numerical}] and is assumed to be statistically independent from the other random parameters, $m_l$ and $f_l$.  We simulated system sizes up to $N=200$ and verified that the qualitative features  of Fig.~\ref{fig:ratio_energy} are robust against variations in specific distributions of the random parameters.     The first order  (red dashed) line in Fig.~\ref{fig:Protocol}  is determined by the position of $(B_z^c(\chi), \chi B_z^c(\chi))$ for a fixed reference state, while $B_z^c(\chi)$ is given by the last intersection of the reference state. \\

\noindent {\bf Acknowledgement}\\
This work was supported by the  NSF grant DMR-2037654.\\

\noindent {\bf Author contributions}\\
All authors participated in developing the theory. H.W., and H.-C.Y. performed the analysis of numerical simulation.  A.K. conceived and supervised the project. All authors contributed significantly to the writing of the manuscript.  \\


\bibliography{IQA.bbl}

\end{document}